# Layered, Tunable Graphene Oxide-Nylon Composite Heterostructures for Wearable Electrocardiogram Sensors

N. G. Hallfors[1], D. Maksimovski[2], I. A. H. Farhat[3], M. Abi Jaoude[1], A. R. Devarajan[1], K. Liao[1], M. Ismail[4], H. Pade[2], R. Y. Adhikari[2], and A. F. Isakovic[2*]

[1] Khalifa University, Abu Dhabi, 127788, United Arab Emirates
[2] Colgate University, Hamilton, NY, 13346, USA
[3] University of Waterloo, Waterloo, ON, Canada
[4] Wayne State University, Detroit, MI, 48202, USA

(*) corresponding author aisakovic@colgate.edu and iregx137@gmail.com

**Abstract** Nanoscale engineered materials combined with wearable wireless technologies can deliver a new level of health monitoring. A reduced graphene oxide-nylon composite material is developed and tested, demonstrating its usefulness as a material for sensors in wearable, long-term electrocardiogram (ECG) monitoring via a comparison to one of the widely used ECG sensors. The structural analysis by scanning electron (SEM) and atomic force microscopy (AFM) shows a limited number of defects on a macroscopic scale. Fourier Transform Infrared (FTIR) and Raman spectroscopy confirm the presence of $rGO_x$, and the ratio of D- and G-features as a function of thickness correlates with the resistance analysis. The negligible effect of the defects and the tunability of electrical and optical properties, together with live ECG data, demonstrate its signal transduction capability.

Electrocardiogram (ECG) sensors measure the electrical voltage produced by depolarization of the cardiac tissue during a heartbeat[1], and accurate ECG relies on the quality of the electrode, as the signal degradation occurs at the skin-electrode interface.[1,2] One frequently used clinical electrode is the wet, gel-type Ag/AgCl adhesive electrode, owing to its good impedance match and signal transduction.[3] These electrodes, however, are only practical with a patient confined to a bed.

With the increasing global prevalence of various chronic conditions, such as Type II Diabetes [4] (see Supporting Online Materials) and associated cardiovascular disease (CVD),[5,6,7,8] there is growing interest in the development of wearable, long-term dry electrodes constant monitoring systems for health parameters.[9,10] Some dry electrodes offer long-term usability but lower signal quality without inconvenient preparation steps for the skin.[11,12] The critical task is to find a suitable electrode material that will be skin-compatible while showing good signal transduction. Achieving these needs is possible with graphene oxide and its derivatives[13, 14, 15]. While many pure graphene production methods involve high-temperature gas-phase deposition techniques,[16,17,18,19] this work relies on an alternative production method practical for large scale-up production. An aqueous dispersion of graphene oxide ($GO_x$) is applied to Nylon fabric in a uniform coating, which is then chemically converted to reduced graphene oxide ($rGO_x$).[20] The $rGO_x$ coating is then characterized via several microscopic, spectroscopic, and electrical measurements, elucidating the parameters behind the usefulness and opening the path for its improvement and large-scale development of ECG electrodes. We relied on a modified Hummer's method[21, 22]. Ultrasonication is used to thoroughly exfoliate the graphite oxide stacks, producing a dispersion of single layer $GO_x$. The dispersion is then purified through several cycles of centrifugation and washing with deionized water, which leads to an aqueous dispersion of $GO_x$. For a fabric substrate, Nylon is selected as a suitable candidate material for its strength, durability under varied chemical conditions, smoothness, and absorbency[23]. A portion of the fabric is cut from a roll, washed, dried, and placed into a volume of a liquid $GO_x$ dispersion. The material is left to soak overnight, ensuring even distribution of liquid over and throughout the entire fabric surface. Next, the fabric is removed, placed on a hydrophobic sheet to prevent excessive runoff of the absorbed dispersion, and left to dry in a convection oven at 90 °C for 2 hours. This results in a piece of fabric with a single coating of $GO_x$, while repeating the process will result in additional layers of $GO_x$. Additional depositions, followed by chemical reduction with hydrogen iodide (HI), are needed to increase the conductivity[24,25,26]. A chemical reduction is conducted in the timed bath of HI. The resulting fabric is highly electrically conductive, strong, durable, and doesn't change properties reported here upon deformation or ambient temperature variations. As such, it is a good candidate for an ECG sensor.[27]

In this report, we will look at some DC and AC electrical properties, optical properties, structural properties, as well as the correlation between the optical and electrical properties. Lastly, we will present sample data for an early version of an ECG sensor. The $rGO_x$ ECG sensor's electrical properties are critical to its functionality. Samples were cut into squares and rectangles of various sizes of potential ECG electrodes (100 – 400 mm$^2$ squares and 50 -200 mm$^2$ elongated rectangles) for electrical characterization, where measurements were performed with Keithley 4200 SCS. Scanning electron microscope (SEM) images were captured on a JOEL JSM-7610. Atomic force microscope images (AFM) were taken on an Asylum MFP-3D in tapping mode. Fourier transform infrared (FTIR) measurements were taken on a Perkin-Elmer Spotlight 200 microscope, and Raman spectra were taken with WITEC. ECG data (AD Instruments PowerLab ECG) were taken from volunteers after proper approval, using $rGO_x$ fabric electrodes at the wrists and neck and simultaneously at the same locations with Ag/AgCl wet electrodes for comparison. Fig. 1 shows the results of the resistance vs $rGO_x$ measurements for the material samples fabricated. The thickness variation is measured with a micro-profilometer after depositing successive layers of $rGO_x$, and after partially masking the preceding coats, leading to a purpose-made step-like structure. Given the unusually large drop-off shown in Fig. 1 (more than a factor of x20) in the resistance value as a function of increasing thickness, we have modeled our layered $rGO_x$ samples in two ways to gain a better insight into processes leading to such a significant change. Model 1 is a circuit model of the layered materials, the simple resistor network shown in Fig. 1.b, with $r_\parallel$ representing the in-plane resistance of a $rGO_x$ monolayer, $r_\perp$ representing the resistance due to out-of-plane conduction, and $R_n$ is the equivalent resistance of the network corresponding to a $rGO_x$ sample with $n$ monolayers. We estimated $n$ for each of our samples as $n = t/\tau + 1 \approx t/\tau$, where $t$ is the measured thickness and $\tau \approx 0.4\ nm$ is the average distance between adjacent rGO$_x$ monolayers. By using the recursive relation (Eq. 1)

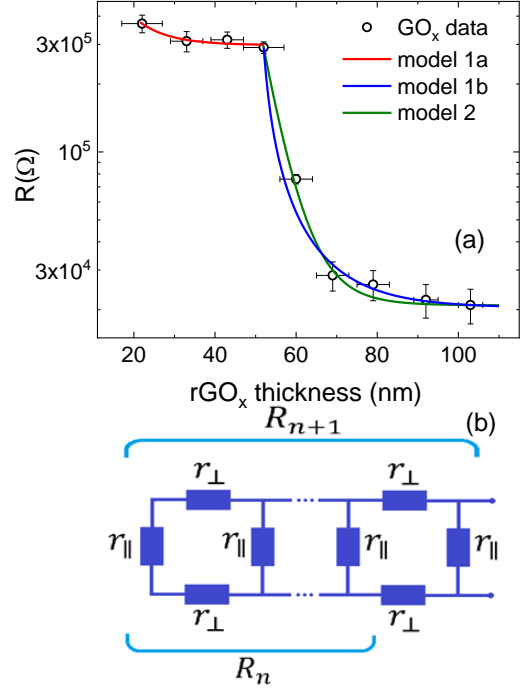

**Figure 1.** *(a) Thickness dependence of the resistance of the multilayered rGO$_x$ films, together with the two models. The model (1a and 1b) is based on resistor network (b). Model 2 relies on the domain structure of rGO$_x$.*

$$R_{n+1} = \frac{(R_n + 2r_\perp)r_\parallel}{R_n + 2r_\perp + r_\parallel} \quad Eq.\ 1$$

we found that $R_n$ models the measured resistances at lower thicknesses best when $r_\parallel = 1.6\ x\ 10^3\ \Omega$, $r_\perp = 2.8\ x\ 10^7\ \Omega$; and models the resistance at higher thicknesses when $r_\parallel = 70\ \Omega$, $r_\perp = 3.0x10^6\ \Omega$. The fact that the transition between these regimes is brought on by order of magnitude decrease in $r_\perp$, and almost two orders of magnitude drop in $r_\parallel$ suggests that the resistance drop might be attributable to a combined in-plane and out-of-plane changes of the transport properties, which in turn is caused by the changes in the in-/out-of-plane bonding as the thickness increases. The second model is motivated by the model of changes in the resistance of multilayered graphene on thickness [29], itself motivated by high-resolution transmission electron microscopy (TEM) studies, which show that the layers of the defect-free crystalline graphene oxide areas are continuous, i.e. can be viewed as continuous arrays of slightly overlapping (sub-)micrometer scale defect-free grains [29]. Due to this, and the defect-free areas being more conductive than the defect regions by at least one order of magnitude, we viewed the structure of rGO$_x$ as being analogous to the grain-structure of CVD graphene - with the defect regions in rGO$_x$ acting similar to the grain boundaries in CVD graphene. At significant chemical reduction, the low defect regions in rGO$_x$ provide a channel for percolation-like transport [30], and are therefore the major contributors to a rGO$_x$ layers' conductivity. Relying on Duong *et al.*[30] empirical relation between the resistance $R$, the thickness and structure factors, we used the expression

$$R = \frac{R_0}{1+(kt)^n} \quad Eq.\ 2$$

where $k$ is a structure factor constant, and $t$ is the thickness. Therefore, we modeled the resistance drop for the sufficiently large thicknesses by an increasing thickness-dependent effective percolation area related to both in-plane and out-of-plane conduction, as shown in the formula above, resulting in Model 2 in Fig. 1a. The resulting fit is shown on Fig. 1a (green curve), and corresponds to the values of $R_0 = 2.09 \cdot 10^4\ \Omega$ and $k = 0.0155\ nm^{-1}$.

In addition to DC analysis in the previous paragraphs, we conducted the Impedance Spectroscopy study, as it has been useful in elucidating the distinction between variations in GO$_x$ and hybid rGO$_x$ properties[31,32]. We used a Gamfry Reference 600+ Impedance Analyzer setup. Beyond this general motivation, we were curious about the frequency dependent changes of layered GO$_x$. Fig. 2 shows the results of this study. Nyquist diagrams in Fig. 2(a) shows characteristic semicircles, which contract inhomogeneously as the rGO$_x$ thickness increases. The decrease in the radia of the Nyquist semicircles is indicative of the changes to the charge transfer mechanism[31], indicated by decreasing time constant. Low thickness samples (below ~ 25 nm) have qualitatively different phase angle curve than other higher thicknesses, so this thickness range is still under study. We can recognize two different regimes low and high frequency ones, when looking at the characteristic inflection points where the phase angle curves "flatten". These regimes could be considered as diffusive and kinetic, and they appear to be of the same order of magnitude as some rGO$_x$ samples in unrelated studies[31,32]. The change in the characteristic frequency is plotted as a function of the sample thickness in Fig. 2(c).

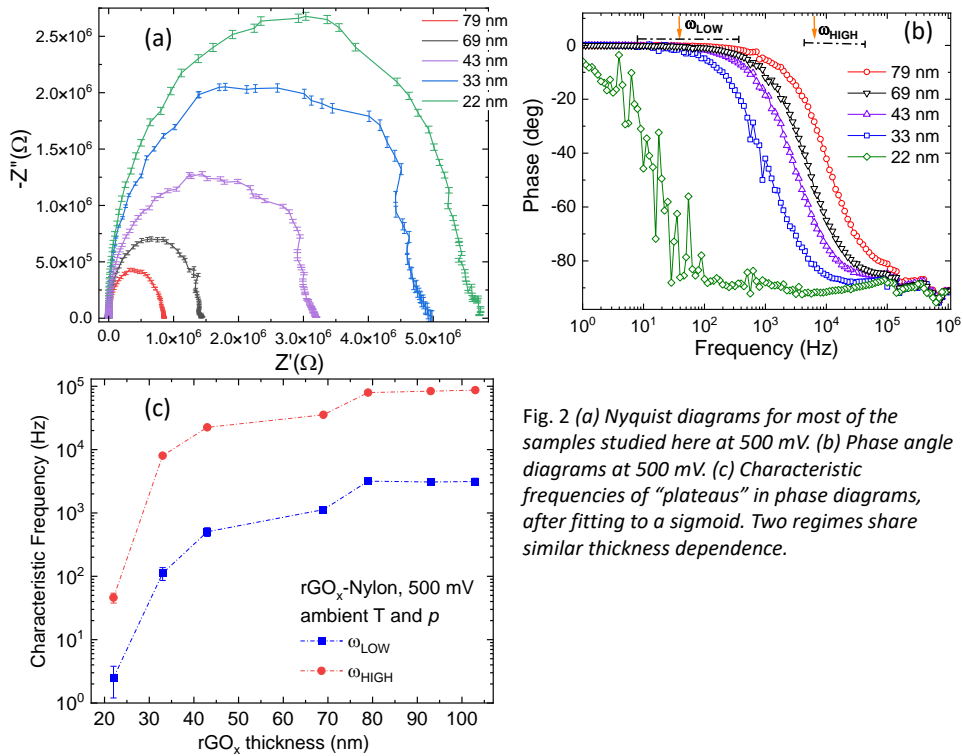

Fig. 2 *(a) Nyquist diagrams for most of the samples studied here at 500 mV. (b) Phase angle diagrams at 500 mV. (c) Characteristic frequencies of "plateaus" in phase diagrams, after fitting to a sigmoid. Two regimes share similar thickness dependence.*

Examining possible dependence between electronic performance and the structural and morphological details is important to understanding the pathways for the improvement of the current material and device level of functionality and integration. Scanning Electron Microscope (SEM) images give us a clear picture of the surface of the composite heterostructure, as well as some insight into the process of surface restructuring as the film thickness increases, which is relevant to the control of the resistivity of the probes as ECG sensors. Figure 3 shows three fabric samples after varying levels of rGO$_x$ thickness (left to right: 0 nm (bare Nylon fabric), 22 and 79 nm) and at increasing magnification. Fig. 3.a shows a fabric which appears uniform and free of defects. At a rGO$_x$ thickness of 22nm, the overall microscopic appearance of the fabric seems to be largely unchanged (Fig. 3.b). Closer examination at high magnification of the bare nylon fiber (Fig. 3.g) shows a series of meso- and nanoscale defects.

Examination of the 22 nm film at high magnification (Fig. 3.h) reveals a drastic reduction of the nanoscale defects, which the rGO$_x$ coating has uniformly covered.

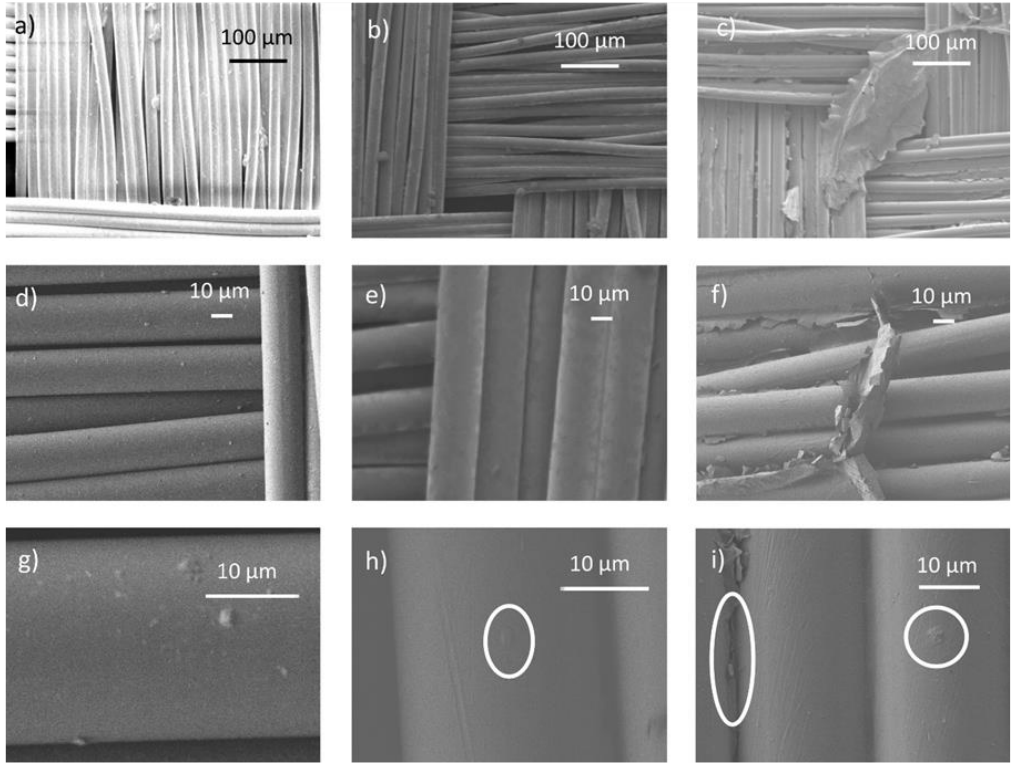

**Figure 3.** *SEM images of (a, d, g) 0nm, (b, e, h) 22nm, and (c, f, i) 79nm rGO$_x$-Nylon heterostructures. The evolution of defects is shown left-to-right and bottom-to-top.*

At slightly lower magnification, however, there appear to be visible discontinuities of the coating (Fig. 3.e). As a result of these discontinuities, the conductivity and signal transduction capability of 22nm rGO$_x$ film both remain relatively low. Examining 79 nm heterostructure (Figs. 3. c, f, and i), we see that while the nanoscale defects from the bare nylon are entirely covered, new defects begin to emerge as a result of the relative excess rGO$_x$ on the fabric surface. The appearance of the significant defects may partially come from a poorly controlled deposition rate, as the substrate is kept in the solution rather than in a vacuum during deposition [32]. Additionally, the regions of poor fiber coverage were observed after a single coating disappeared, and electrical conductivity has risen significantly due to the absence of observable voids. With subsequent re-coatings, excess rGO$_x$ can be observed physically connecting adjacent fibers (Fig. 3.i). While the electrical conductivity has drastically increased with multiple deposition steps, the more significant defects become a concern as they may become nucleation points from which the structure grows in an unintended manner (i.e., not surrounding the core nylon matrix and/or not coating it uniformly), and therefore may lead to less reliable material. Relatively large-scale robustness of the ECG sensors proposed here, despite the presence of some defects, points toward (a) a pathway for future improvement through defects' engineering, and (b) the current level of applicability of rGO$_x$ in ECG and related sensors, motivating this work beyond optical[33] and biochemical sensors[34]. Atomic force microscope (AFM) images were obtained to provide additional micro–and nano–characteristics at the composite fabric surface. As in the SEM images, nanoscale defects are noticeable in the bare fabric (Fig. 4.a). These defects start to become smoothed over by a thin film of rGO$_x$ (Fig. 4.b) and are subsequently replaced by more significant microscale defects in the rGO$_x$ film (Fig. 4. c,d) as additional rGO$_x$ deposition occurs. Figs. 4a-d represents 6x6 µm$^2$ scans over single fibers. The broad range of changes in the relative variation of heights appears to be 1.2 to 5.3 nm. In the past work, [34] GO nanosheets were found to have a thickness of the order of 1.0 nm, and rGO$_x$ nanosheets are typically 0.3 - 0.6 nm. These values can be obtained by AFM, and the change in thickness could likely be attributed to the removal of the various out-of-plane oxide groups. Measurements with scanning tunneling microscopy (STM) support these findings.[35] From our observations of defects in the 1-5 nm range, we have good reason to believe that these defects arise from a small number of rGO$_x$ sheets (likely fewer than 5) either folding back on themselves or randomly overlapping.[36] Alternatively, these defects could be the result of unreduced (or, more precisely, less than "average reduced") sheets remaining randomly dispersed within the film.[37] Understanding of the structure of rGO$_x$ films that goes beyond morphological studies (such as AFM and SEM) is needed, and we offer the results of Fourier Transform Infrared (FTIR) spectroscopy in the next section.

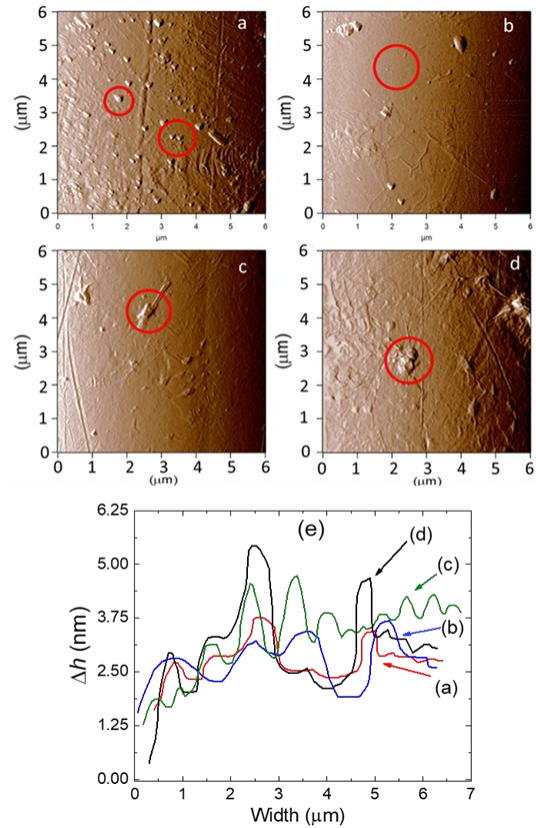

**Figure 4.** *(a) - (d) show AFM images of 0 nm (a), 22 nm (b), 43 nm (c) and 79 nm rGO$_x$ film respectively, on nylon fiber. Nanoscale defects in the bare nylon (a) are covered up with the initial rGO$_x$ coating (b) and two subsequent coatings in (c) and (d). (e) Topographical scans extracted from AFM data show surface defect height range is from 0.6 nm and 5.2 nm, and defects lateral size is limited to approximately 300 nm.*

In addition to the correlation between morphology and structure, performing FTIR is important as we look to understand the role of different functional groups in the process of stepwise increase of film thickness, which itself is necessary to obtain proper electrical and optical characteristics of rGO$_x$. The FTIR spectra of samples in Fig. 5(a) showed many of the typical vibrational peaks of graphene oxide and its derivatives, such as 980 cm$^{-1}$, (sp$^2$ C-H stretching mode), 1016 cm$^{-1}$ (alkoxy stretch at C-O moiety), ~ 1340 cm$^{-1}$ (D-feature, associated with the C-O-C (epoxide) group, involving a phonon-defect interaction in the sp$^2$ graphitic structure), 1504 cm$^{-1}$ (sp$^2$ hybridized C=C bond), 1574 cm$^{-1}$ (G-feature, first-order Raman scattering as a single E2g phonon optical vibration in the graphitic structure), and 1711 cm$^{-1}$ (carboxyl stretching mode, likely picking up contributions from both COOH and C=O) and several others, as shown in Fig. 5 (a).[38,39,40] As the atomic content in the film varies with the increasing thickness, some changes in the FTIR spectral width and centerline position are expected. Most features have small change in the peak position $\Delta\omega_C$, about 1-3 cm$^{-1}$, Fig. 5 (c) through (h). The findings indicate that different bonding mechanisms contribute differently to structural properties of rGO$_x$ with the increasing thickness. Overall, the change in absorbance is smaller with the increasing thickness, as one would expect for the thickness-limited process. We note the small scale of the changes and the near-linear behavior, in contrast to the electrical properties. A more significant change is seen in two relevant features: sp$^2$ C-H bend at 980 cm$^{-1}$ ($\Delta\omega_C$ = 24 cm$^{-1}$) and epoxide C-O-C asymmetric stretch near 1240 cm$^{-1}$ ($\Delta\omega_C$ = 10 cm$^{-1}$). Based on available GO$_x$ literature, and our study, we propose that these two features, together with the carboxyle moieties at 1711 cm$^{-1}$ discussed above, are most responsible for the defect creation discussed in SEM and AFM sections, and therefore point towards likely development path of further improvement of this material for sensor through the tuning of the wet chemistry process. To this end, we conducted Raman spectroscopy scans, and analyzed the ratio of the D- and G- spectral features ($I_D/I_G$), as a function of the film thickness, shown in the bottom panel of Fig. 6, leading to interesting observations.

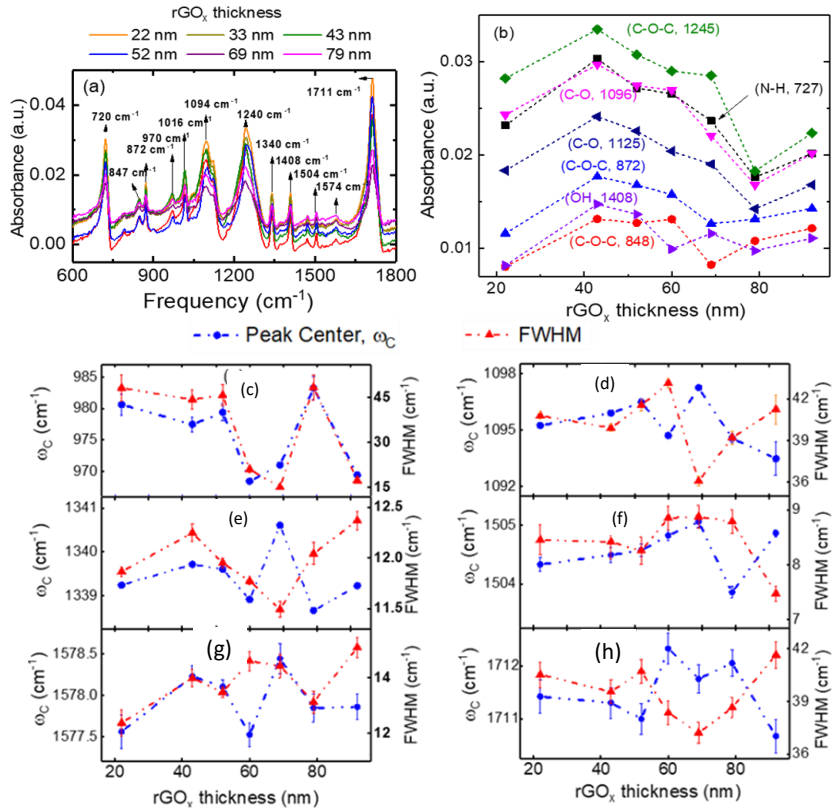

**Figure 5**. *(a) ATR FTIR data in a mid-infrared range shows a number of structurally relevant features of rGO$_x$. (b) change in absorbance as a function of film thickness for most relevant rGO$_x$ features. Small changes in FWHM and the center peak for six of the spectral features from panel (a), indicating the quality of the heterostructure.*

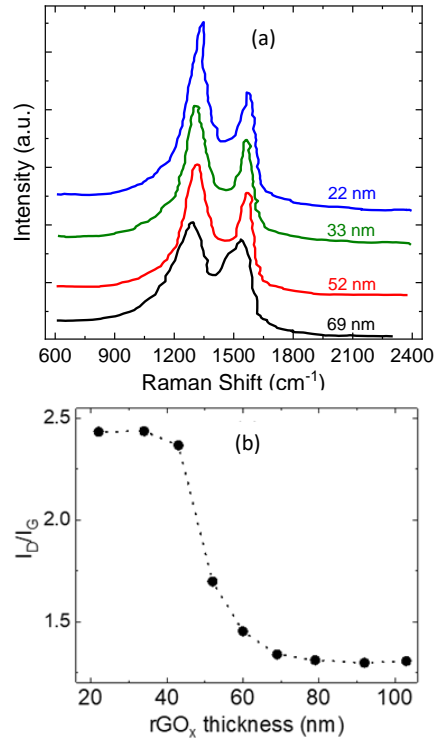

**Figure 6**. *(a) Raman spectra for samples in this study as thickness changes (b) A relative intensity of D and G features, with the uncertainties the size of the symbol or smaller.*

First, the overall scale of change in $I_D/I_G$ is comparable to most nanoscale (1-10 nm) thick $GO_x$ sheets, and it is relatively small (between a factor of 1 and 2.5). Second, the spread of FWHM values is larger for the G- feature than for the D-feature, confirming the specific role of the C-O-C bond in the multilayered growth of $GO_x$[39,40,41,42], suggesting the path toward improving the homogeneity of the film is through more careful annealing of defects and different variants of Hummer's method [43]. Lastly, we note the apparent similarities between the thickness dependences of resistance and the $I_D/I_G$ ratio, with the caveat that the former one is on a different quantitative scale. This indicates a correlation between some electronic and some optical and structural properties, related to the role disorder plays in $rGO_x$[44].

Specifically, broader D-band at lower thicknesses reflects the change in the $sp^2$ domain size, while the shift in G-band indicates changes from dominantly $sp^2$ to some presence of $sp^3$ bonds. This is also reflected in the changes of $R_{perp}$ and $R_{\|}$ changes in the model in Fig. 1.

As a test of an intended application, rGO$_x$ dry electrodes performed comparably or better as the standard wet electrodes for ECG signals. The composite fabric electrode ($t$ = 103 nm) returned an ECG signal with clearly identifiable PQRST peaks (Fig. 7a, inset) and displayed higher ECG signal amplitude and lower noise compared to the wet electrode. (Fig. 7a, d). The period between the T and P waves in ECG signal with essentially no muscle activity is referred to as the isoelectric region, and it is used as an indicator of the overall noise in, and therefore quality of, the ECG signal.[45, 46] Fig. 7(d) shows the isoelectric noise vs. activity for ECG data recorded simultaneously on both electrodes, with a distinct difference in the behavior of the two electrodes, in favor of the lower point spread with the rGO$_x$ electrode. Both, higher amplitude and lower noise are encouraging and warrant both development of sophisticated wearable electronics and clinical trials, especially as the level of noise is a key measure of the signal quality.[41]

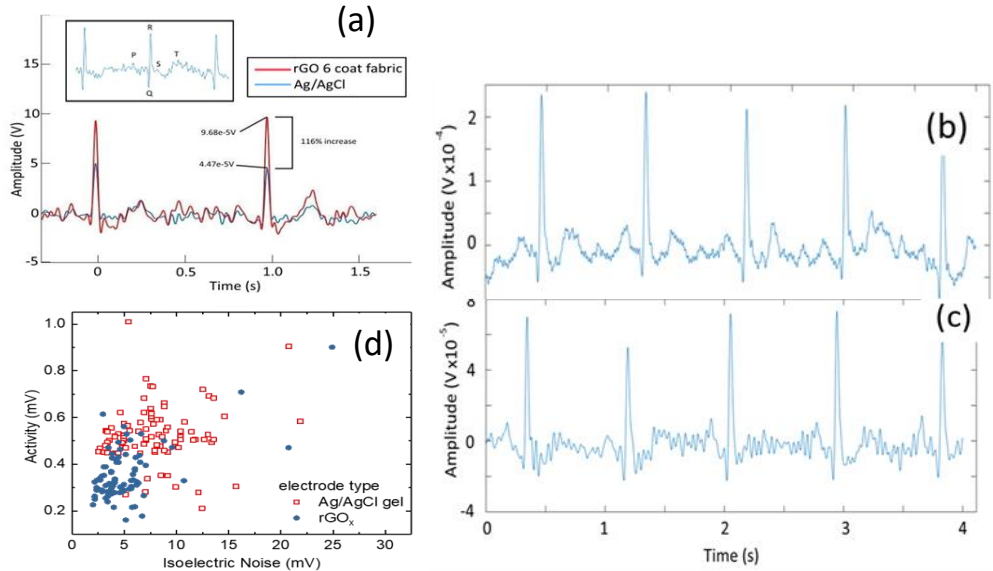

**Figure 7.** *(a) Overlay of live ECG data collected simultaneously with Ag/AgCl gel-type electrodes and dry rGO$_x$ electrodes (t = 103 nm), with characteristic PQRST waveform **(inset)** clearly identifiable in both signals. (b) and (c) ECG signal from two different size rGO$_x$ electrodes, 1cm$^2$ and 4 cm$^2$, respectively. (d) Activity vs isoelectric noise for conventional and dry rGO$_x$ electrodes.*

Potentially equally relevant for the applications, the signal increases with the area of contact between the $rGO_x$ electrode and human skin, as shown in Fig. 7 (b, c). Finally, signal-to-noise ratio was 28.3 for $rGO_x$ electrode and 24.9 for commercial one. Data were taken with $rGO_x$ and gel-type electrodes on the same volunteers, under the same environmental and body conditions. Hundreds of ECG runs can be collected without significant changes to the signal S/N ratio. Additional studies, including durability, are necessary to see whether the devices based on such electrodes completely replace the current ones.

In summary: (a) DC and AC electric properties show application promise in ~ 100 nm thickness range; (b) electrical properties thickness dependence correlates with thickness dependence of the characteristic Raman peaks' ratio; (c) a number of IR spectral features are slowly varying or approximately constant with thickness; (d) SEM and AFM images identify the texture of the $rGO_x$ coating and promise for the future optimization needed for the build-up of on-chip, integrable wearable electronics for IoT healthcare, (e) three types of ECG data demonstrate the applicability of this material as a potential ECG sensor. One can expect more than one type of sensor could be developed based on this material. The multilayer $rGO_x$ heterostructures are highly tolerant of the presence of defects, as seen in AFM and SEM studies. Future study of this material will involve more sophisticated fabrication techniques to achieve full in-plane and out-of-plane characterization of transport and bonding properties. Based on the overall electrical, structural, and spectroscopic data, it is clear that one can design a multiprobe $rGO_x$ ECG sensor where the range of physical properties of the component probes spans the relevant conditions of the patient's skin and overall health, as well as environmental conditions, such as temperature and humidity. These environmental and patient condition variations can contribute to changes of the overall impedance of the skin, and yet this material remains a viable ECG sensor material candidate.

As a matter of perspective and future work, the $rGO_x$ electrode outperformed the industry standard gel-type Ag/AgCl electrode in terms of signal amplitude and noise level for ECG data collected from the wrist and the neck. Additional work is necessary to understand the bonding of the initial layer(s) of $rGO_x$ to Nylon, and while some of the data reported here might be useful for such considerations, the think additional techniques and efforts are needed to fully elucidate such bonding. Data were taken with $rGO_x$ and gel-type electrodes on the same volunteers, under the same environmental and body conditions. Hundreds of ECG runs can be collected without major changes to the S/N ratio. Additional studies, including durability and long-term environmental effects, are necessary to see whether the devices based on such electrodes completely replace the current ones. It is essential to test $rGO_x$ ECG electrodes for their multi-day effectiveness under extreme outdoor conditions of temperature and humidity, with the initial tests being promising towards satisfying these criteria.

**Acknowledgments:** We acknowledge the assistance of Dr. Y. A. Samad in the early stages of $rGO_x$ growth and Dr. Ahsan Khandoker for the use of the Electrophysiology Laboratory at KU.

**Funding:** This work was supported by the Mubadala-SRC through the 2013-HJ-2440 grant and, in part, through the SRC 2011-KJ-2190. KL and AFI acknowledge funding from 2013-KUIRF-L2. AFI acknowledges the support and hospitality of Cornell University Cornell CNF, funded through NSF. DM, HP, RYA, and AFI acknowledge Summer Research Support at Colgate University.

**Data Availability:** Data are available upon reasonable request.

**Supplementary Materials:** Accompanying file contains additional information about this work.